\documentclass[12pt]{iopart}

\usepackage{graphicx}
\usepackage{amssymb}

\begin{document}

\title[Limits on Quantum Probability Rule by no-Signaling Principle]{Limits on Quantum Probability Rule by no-Signaling Principle}

\author{Yeong Deok Han$^1$ and Taeseung Choi$^2$}

\address{$^1$Department of Game Contents, Woosuk University, Wanju, Cheonbuk, 565-701, Korea }
\address{$^2$Division of General Education, Seoul Women's University, Seoul 139-774, Republic of Korea }
\ead{\mailto{tschoi@swu.ac.kr}}


\begin{abstract}
     We have studied the possibility of post-quantum theories more nonlocal than the 
     (standard) quantum theory using
the modification of the quantum probability rule under the no-signaling condition.
For this purpose we have considered the situation that two spacelike separate parties Alice and Bob share
 an entangled two qubit system. 
 We have modified the quantum probability rule as small as possible such that 
 the first local measurements are governed by the usual Born rule 
 and the second measurement by the modified quantum probability rule. 
 We have shown that only the maximally entangled states can have higher nonlocality than the 
 quantum upper bound while satisfying the no-signaling condition.
This fact could be a partial explanation for why the nonlocality of the quantum theory is limited. 
As a by-product we have found the systematic way to obtain a variety of nonlocal boxes.
\end{abstract}
\maketitle

\section{Introduction}
Quantum mechanics is nonlocal in the sense that no local hidden variable theories can simulate quantum mechanical correlations \cite{Bell}.
The nonlocality of quantum mechanics can be demonstrated by a violation of inequalities on measurable correlations.
In the Clauser-Horne-Shimony-Holt (CHSH) inequality the degree of violation is represented by the value of
the CHSH parameter $\mathcal{B}$ \cite{Clauser}. 
The quantum upper bound of $\mathcal{B}$, the degree of nonlocality, is known as Tsirelson's bound
$\mathcal{B}=2\sqrt{2}$ \cite{Tsirelson}.
The nonlocality of quantum mechanics, however, is not maximum as found by Popescu and Rohrlich \cite{Popescue1}.
They derived ``superquantum'' correlation by using two axioms of nonlocality and relativistic causality.
The maximum value $4$ of $\mathcal{B}$ is obtained for Popescu-Rohrlich (PR) nonlocal boxes \cite{Popescue1}.
The nonlocality of the PR box is limited only by the no-signaling principle whereby no information can be transferred
faster than the speed of light.
Therefore we naturally ask a question on the possibility of the existence of modified quantum theories, so called post-quantum theories, which are
more nonlocal than the quantum theory and limited only by the no-signaling principle.

The properties of post-quantum theories are studied by several authors in an information theoretic view
\cite{Dam, Barrett, Brassard1, Linden1, Navascues, Brunner, Brunner1, Pawlowski,Cavalcanti, Gallega, Yang, Dahlsten}. 
As far as we know, however,
there is no approach to find post-quantum theories by extending the quantum theory directly.
The main postulates of the quantum theory are consisted of two parts.
One is for a system and a time evolution of the system, which is
based on a ray in a Hilbert space and a unitary evolution operator \cite{Weinberg1}.
The other is for a measurement interpretation. Hilbert space based description of physical states
is natural for a linear quantum mechanics.
Projective measurement postulate, however, is justified only by experiments. 
The study on the possibility of higher-path interferences than two-path interference which corresponds to
the generalization of the Born rule has been done both in theories and in experiments \cite{Sorkin, Ududec, Sinha}.
Therefore modifications of measurement postulates
are good candidates for extending the quantum mechanics to more nonlocal post-quantum mechanics.

In the present paper, we treat the problem to find the possibility of post-quantum theories 
by asking a question of simulating the PR box by modifications of quantum probability postulate 
for quantum measurements.
The standard quantum probability rule known as the Born rule is considered to be restricted by Gleason's theorem \cite{Gleason}.
 Gleason's theorem states that the only consistent probability assignment for all projection operators on Hilbert 
spaces dimension at least three must follow the standard quantum probability rule.
In our study we will consider an entangled state which lives in 4-dimensional Hilbert space, therefore,
it seems the modification of the Born rule is prohibited by Gleason's theorem in simple-minded consideration, however,
in our bipartite system only the local projection operators for each party are allowed. On the other hand, to prove 
Gleason's theorem all possible projection operators including the global operators must be involved.
Moreover, the noncontextuality of probability induced by Gleason's propositions is lack of physical basis.
We impose the no-signaling condition instead of the noncontextuality 
which is compatible with the relativity on the Hilbert space 
formalism of the physical states.
We will organize this paper as follows. We will explain the formalism of the modification of the Born rule in section 2.
In section 3 we will show the joint probability distribution of the PR box can be simulated by 
the maximally entangled state under the modified Born rule. In section 4 we will study the nonlocality of 
other nonlocal boxes which can be obtained systematically by changing the observables and the quantum probability rule.
In section 5 we will summarize our results and discuss the implications of the results.


\section{The formalism for the modification of the Born rule}

 
  In our study,
we assume that a physical state is described by a vector $|\psi\rangle$ in a Hilbert space with usual norm
$| \psi |=\langle \psi |\psi\rangle^{1/2}$.
A physical observable is represented by a linear and Hermitian operator $\mathcal{A}$.
An observable $\mathcal{A}$ has real eigenvalues $\{ a_1, \cdots, a_d\}$ and mutually orthonormal eigenvectors
$\{ |\phi_1\rangle, \cdots, |\phi_d\rangle\}$, where $d$ is the dimension of the Hilbert space.
Physical observables satisfy the following measurement postulates: i) an outcome of measurement is always an eigenvalue of
$\mathcal{A}$. ii) The probability of an outcome $a_k$ for an initial state $|\psi\rangle$ is obtained
with $f(a_k)=|\langle \phi_k|\psi\rangle|^2$.
iii) The quantum state after the measurement that gives the outcome $a_k$
reduces to the corresponding eigenstate $|\phi_k\rangle$.
Where $f(a_k)$ is the function
which assigns the probability for the outcome $a_k$.
The postulate ii) is the quantum probability rule, which is called the Born rule.
The Born rule states that $\langle \mathcal{A} \rangle = \mbox{tr} \left( \rho \mathcal{A}\right)$
in a density matrix formulation.
Where $\langle \mathcal{A} \rangle$ is the expectation value of the observable $\mathcal{A}$
and $\rho$ is a density operator $|\psi\rangle \langle \psi|$.

Let us analyze the modification of the quantum theory.
For this purpose, let us consider the possibility of changing the postulate ii) of the quantum measurement.
According to Gleason's theorem probability assignment for each vector
of an orthonormal basis $\{ |\phi_1\rangle, \cdots, |\phi_d\rangle\}$ associated with
the state vector $|\psi\rangle$ must be $f(a_k)=|\langle \phi_k |\psi\rangle|^2$
under a non-contextual condition, in dimensions at least three. 
In our study we will consider entangled states of two qubits which live in 4-dimensional Hilbert space,
hence, it seems that the modification of the Born rule is restricted by Gleason's theorem. 
This, however, is not the case in our physical situation. 
Suppose Alice and Bob are in spacelike separated positions and share an entangled state.
We impose a physical requirement that every projective measurement by Alice and Bob must be local.
This requirement is compatible with the theory of relativity.
Therefore only product type local projectors such as $\mathcal{P}_A \otimes \mathcal{P}_B$ are allowed,
where subscripts $A$ and $B$ represent Alice and Bob respectively.
For simplicity we assume that Alice measures first and Bob measures second so that 
we only involve $\mathcal{P}_{A1} \otimes \mathcal{P}_{B2}$ in our theory.
In Gleason's proposition, however, any projection operators are allowed so our situation is
different from the proposition of Gleason.
We impose only the condition of the no-signaling and local observables 
as a physical requirement on the Hilbert space 
formalism of the physical state. The condition of the no-signaling and local observables 
is compatible with the theory of relativity which restricts faster-than-light signaling.

We prefer that the modification from the quantum theory is as small as possible.
Therefore we will use the standard Born rule for the quantum measurements of Alice and
modify the Born rule only for the quantum probability rule of Bob. 
 First we will explain the modification of the Born rule in two-dimensional Hilbert space for simplicity.
For a linear and Hermitian operator $\mathcal{A}$ with
two orthonormal eigenstates $|\phi_0 \rangle$ and $|\phi_1 \rangle$,
when a system in a state $|\psi\rangle=\alpha_0 |\phi_0\rangle + \alpha_1 |\phi_1\rangle$ is measured,
measurement probabilities to get outcomes $a_0$ and $a_1$ are modified from the Born rule to more general functions 
such as
\begin{eqnarray}
\label{eq:modify}
f(a_0)= F_0 (\alpha_0, \alpha_1) \mbox{ and }
f(a_1)= F_1 (\alpha_0, \alpha_1),
\end{eqnarray}
respectively.
Where $F_0 (\alpha_0, \alpha_1)$ and $F_1 (\alpha_0, \alpha_1)$ are some non-negative functions which
satisfy the following obvious conditions;
$ F_0 (\alpha_0, \alpha_1) + F_1 (\alpha_0, \alpha_1) = 1$ (normalization), 
$ F_0 (\alpha_0, \alpha_1) = F_1 (\alpha_1, \alpha_0)$ (bases relabeling invariance), 
$ F_0 (\alpha_0, \alpha_1) = F_0 (|\alpha_0|, |\alpha_1|)$ (phase redefinition invariance), and 
$ F_0 (\alpha_0, \alpha_1) = F_0 (s \alpha_0, s \alpha_1)$ (state normalization).
The Born rule corresponds to $F_0(\alpha_0, \alpha_1)= |\alpha_0|^2 /(|\alpha_0|^2 +|\alpha_1|^2) $.
We will call this modified assignment for measurement probability
as 'modified quantum probability rule'.
After a measurement the state $|\psi\rangle$ is projected either to $|\phi_1\rangle$ or $|\phi_2\rangle$
as the same in the quantum mechanics.



\section{The joint probability distributions under the modified Born rule}

We will first apply this modified quantum probability rule to study the possibility of simulating the PR box.

\subsection{PR box}
Suppose Alice and Bob share two boxes with inputs and outputs in their locations. 
 Let $x, y \in\{0, 1\}$ be inputs of Alice and Bob, and $a, b \in\{0, 1\}$ be outputs for Alice and Bob respectively.
The PR box is defined by these two boxes with inputs and outputs having the following hypothetical correlations:
\begin{eqnarray}
a \oplus b=xy,
\end{eqnarray}
where $\oplus$ denotes an addition modulo $2$.
For the above correlations the joint probability $P(a,b|x,y)=1/2$ and for the other cases $P(a,b|x,y)=0$.
We have used a normalization in which the sum of $4$ possible outcomes to the given inputs
is unity.
It is known that the PR box correlations are nonlocal and
cannot be reproduced classically and quantum mechanically \cite{Popescue1}.

\begin{figure}
\centering
\includegraphics[scale=0.3]{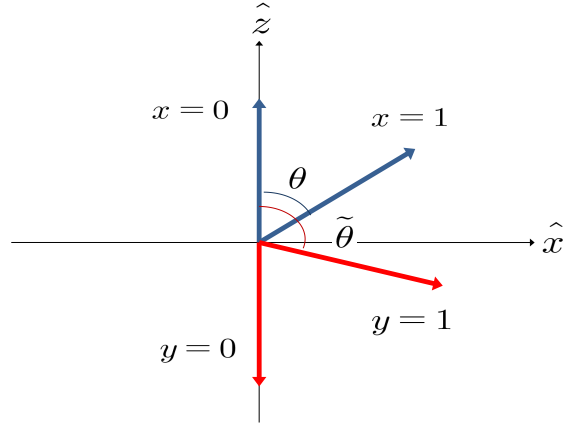}\\
\caption{(Color online) The schematic diagram for measurement axis of Alice and Bob.}
\label{fig:1}
\end{figure}

\subsection{Physical situations}
Our purpose is to study the possibility of making the hypothetical correlations of the PR box 
between Alice and Bob by quantum states with the modified quantum probability rule for Bob's measurement.
  Now suppose that the following entangled state is shared between Alice and Bob,
\begin{eqnarray}
\label{eq:state}
|\psi\rangle_{AB} = \alpha|\uparrow\rangle_A |\downarrow\rangle_B + \beta
|\downarrow\rangle_A |\uparrow\rangle_B,~~~
|\alpha|^2 + |\beta|^2 = 1,   
\end{eqnarray}
where subscripts $A$ and $B$ represent Alice and Bob respectively as before. 
$|\updownarrow\rangle $ are eigenstates
of Pauli operator $\sigma_z$.
Suppose the outcomes of $|\uparrow\rangle$ and $|\downarrow\rangle$ are $0$ and $1$ respectively
for $\sigma_z$ measurement to use the same notation as the outputs of the PR box.
Alice and Bob each have two different axis of measurement such as in Fig. 1,
where the arrowhead represents the
measurement direction which gives outcome $0$. The outcome corresponding to the opposite
direction is $1$. 
Where the angle $\theta$ is measured from the positive $z$-axis to the positive $x$-axis.
Alice measures along one of two directions $z$ and $\theta$ which correspond to one of two inputs $x=0$ and $x=1$.
The relations between eigenstates of $\sigma_z$ and $\sigma_\theta$ are as follows
\begin{eqnarray}
\label{Eq:StateRelation}
|\uparrow\rangle_A = \cos{\frac{\theta}{2}} |\theta \rangle_A + \sin{\frac{\theta}{2}} |\theta_\perp \rangle_A,~~ 
|\downarrow\rangle_A = -\sin{\frac{\theta}{2}} |\theta \rangle_A + \cos{\frac{\theta}{2}} |\theta_\perp \rangle_A.
\end{eqnarray}
Where $|\theta \rangle$ and $|\theta_\perp \rangle$ are two orthonormal 
eigenstates of $\sigma_\theta$ corresponding to outcomes $0$ and $1$ respectively.
Bob takes measurements along either $-z$ ($y=0$) or $\tilde{\theta}$ ($y=1$) directions.
 Then relationships similar to Eq. (\ref{Eq:StateRelation}) hold between eigenstates $\{|\uparrow\rangle_B, |\downarrow\rangle_B\}$ and
$\{|\tilde{\theta}\rangle_B, |\tilde{\theta}_\perp\rangle_B\}$.
We will explain calculations of $P(a,b|x,y)$ for the initial state $|\psi\rangle_{AB}$ with some typical inputs
and outcomes under the 'modified quantum probability rule' for Bob's measurement and summarize all results.

We assume Alice and Bob are not moving relative to each other in order not to make any
complexity related to simultaneity. 
Under our proposition the measurement of Alice is taken first and the quantum probability rule 
for Alice is governed by the usual Born rule. 
After the measurement of Alice the projected state becomes the pure separable state and so 
Bob's measurement is taken on the pure state in his two-dimensional Hilbert space.
The probability of Bob's measurement will follow the modified quantum probability rule.

\subsection{Calculations for the joint probabilities}
Let us first consider a calculation of $P(0,0|0,0)$. In our case the $P(0,0|0,0)$ is calculated by
the probability of outcome $b=0$ in Bob's $y=0$ measurement under the condition that Alice's outcome $a=0$ is given
in her $x=0$ measurement.
The outcome $a=0$ for $x=0$
is given only by $ \alpha|\uparrow\rangle_A |\downarrow\rangle_B $ term of the initial state in Eq. (\ref{eq:state}).
The projection probability from the initial state to $|\uparrow\rangle_A |\downarrow\rangle_B$ state is
$|\alpha|^2$ since this projection is taken by Alice's measurement.
After the measurement of Alice, Bob's state projected onto the state $|\downarrow\rangle_B$
because of the entanglement.
This state gives outcome $b=0$ for the measurement $y=0$ with probability $1$.
Therefore $P(0,0|0,0)=|\alpha|^2$. For measurement axis $x=0$ and $y=0$ there are no new results related
to the modified quantum probability rule for Bob's measurement
 since the state of Bob after Alice's measurement is not a superposition
of two outcome states.  That is, $f(a_0)=1$ for $f(a_1)=0$ and vice versa in Eq. (\ref{eq:modify}).

Next consider $y=1$ case, for example, $P(0,0|0,1)$.
For $a=0$ and $x=0$ the state of Bob is projected onto $|\downarrow\rangle_B$ with probability
$|\alpha|^2$ as before.
Since Bob will take the $y=1$ measurement, the state of Bob must be rewritten as
\begin{eqnarray}
|\downarrow\rangle_B = -\sin{\frac{\tilde{\theta}}{2}}|\tilde{\theta}\rangle_B +
\cos{\frac{\tilde{\theta}}{2}}|\tilde{\theta}_\perp\rangle_B,
\end{eqnarray}
by using the eigenstates of the $y=1$ measurement, $|\tilde{\theta}\rangle_B$ and $|\tilde{\theta}_\perp\rangle_B $
corresponding to outcomes $0$ and $1$ respectively.
The outcome probability will be obtained by our 'modified quantum probability
rule' such that the probability of outcome $b=0$ for $y=1$ measurement will be
$ F_0(-\sin{\frac{\tilde{\theta}}{2}}, \cos{\frac{\tilde{\theta}}{2}}) $.
Therefore
\begin{eqnarray}
 P(0,0|0,1)= |\alpha|^2 \space F_0(-\sin{\frac{\tilde{\theta}}{2}}, \cos{\frac{\tilde{\theta}}{2}}).
\end{eqnarray}

One of other nontrivial cases is $P(0,0|1,0)$. Here Bob's measurement looks the same as that in the first case 
of $P(0,0|0,0)$, however, the nontriviality appears because of the different measurement of Alice. 
To get this joint probability Alice must take her measurement with $\sigma_\theta$ first.
In this case the initial state must be rewritten by using the eigenstates of $x=1$ measurement such as
\begin{eqnarray}
|\psi\rangle_{AB} &=& |\theta\rangle_A \left\{ \alpha\cos{\frac{\theta}{2}}|\downarrow\rangle_B -\beta \sin{\frac{\theta}{2}}|\uparrow \rangle_B \right\} \\ \nonumber
&+& |\theta_\perp\rangle_A \left\{ \alpha\sin{\frac{\theta}{2}}|\downarrow\rangle_B +\beta \cos{\frac{\theta}{2}}|\uparrow \rangle_B \right\}.
\end{eqnarray}
Then the probability of outcome $a=0$ for $x=1$ is $|\alpha \cos{\frac{\theta}{2}}|^2+ |\beta \sin{\frac{\theta}{2}}|^2$
according to the Born rule.
After the measurement of Alice the state of Bob is projected onto the following normalized state,
\begin{eqnarray}
\frac{1}{\sqrt{|\alpha|^2 \cos^2{\frac{\theta}{2}} + |\beta|^2 \sin^2{\frac{\theta}{2}}}}
\left\{ \alpha\cos{\frac{\theta}{2}}|\downarrow\rangle_B -\beta \sin{\frac{\theta}{2}}|\uparrow \rangle_B \right\}.
\end{eqnarray}
The probability of outcome $b=0$ for $y=0$ measurement will be
$ F_0(\alpha \cos{\frac{\theta}{2}}, -\beta \sin{\frac{\theta}{2}}) $
according to the modified quantum probability rule.
Therefore the result becomes
\begin{eqnarray}
 P(0,0|1,0)= \left( \left| \alpha \cos{\frac{\theta}{2}} \right|^2 + \left| \beta \sin{\frac{\theta}{2}} \right|^2 \right)   F_0( \alpha \cos{\frac{\theta}{2}}, -\beta \sin{\frac{\theta}{2}}) .
\end{eqnarray}
Note that only local measurement of Bob follows our new quantum probability rule 
and the others are governed by the standard Born rule.

The summary of all results for the probability distributions $P(a,b|x,y)$ are as follows:
\\
1. For $x=0$ and $y=0$ measurements,
\begin{eqnarray}
P(0,0|0,0) &=& |\alpha|^2, ~~~~P(1,0|0,0)=0, \\ \nonumber
P(0,1|0,0) &=& 0, ~~~~~~~P(1,1|0,0)=|\beta|^2.
\end{eqnarray}
The total probability $P(x=0;y=0)$ which is the sum of all possible outcomes for inputs $x=0$ and $y=0$,
 i.e., $P(0,0|0,0)+P(1,0|0,0)+P(0,1|0,0)+P(1,1|0,0)$ is $1$ as expected.
\\
2. For $x=1$ and $y=0$ measurements,
\begin{eqnarray}
\label{Eq:PD2} \nonumber
P(0,0|1,0)&=& \mathcal{A}_1
  F_0(\alpha \cos{\frac{\theta}{2}}, \beta \sin{\frac{\theta}{2}}) , \\ 
P(1,0|1,0) &=& \mathcal{A}_2
  F_0(\alpha \sin{\frac{\theta}{2}}, \beta \cos{\frac{\theta}{2}}), \\ \nonumber
P(0,1|1,0) &=& \mathcal{A}_1
  F_0(\beta \sin{\frac{\theta}{2}}, \alpha \cos{\frac{\theta}{2}}),  \\ \nonumber
P(1,1|1,0) &=& \mathcal{A}_2                             
  F_0(\beta \cos{\frac{\theta}{2}}, \alpha \sin{\frac{\theta}{2}}) ,
\end{eqnarray}
where $\mathcal{A}_1 \equiv |\alpha|^2 \cos^2{\frac{\theta}{2}}+|\beta|^2 \sin^2{\frac{\theta}{2}}$ and
$\mathcal{A}_2 \equiv |\alpha|^2 \sin^2{\frac{\theta}{2}}+|\beta|^2 \cos^2{\frac{\theta}{2}}$.
\\
3. For $x=0$ and $y=1$ measurement,
\begin{eqnarray}\nonumber
P(0,0|0,1)&=& |\alpha|^2
  F_0(\sin{\frac{\tilde{\theta}}{2}}, \cos{\frac{\tilde{\theta}}{2}})   \\
P(1,0|0,1) &=& |\beta|^2
  F_0(\cos{\frac{\tilde{\theta}}{2}}, \sin{\frac{\tilde{\theta}}{2}})   \\ \nonumber
P(0,1|0,1) &=&|\alpha|^2
  F_0(\cos{\frac{\tilde{\theta}}{2}}, \sin{\frac{\tilde{\theta}}{2}})   \\ \nonumber
P(1,1|0,1) &=& |\beta|^2
  F_0(\sin{\frac{\tilde{\theta}}{2}}, \cos{\frac{\tilde{\theta}}{2}})  .
\end{eqnarray}
4. For $x=1$ and $y=1$ measurements,
\begin{eqnarray} \nonumber
P(0,0|1,1) &=& \mathcal{A}_1  F_0(\mathcal{C}_1, \mathcal{C}_2),  \\
P(1,0|1,1) &=& \mathcal{A}_2 F_0(\mathcal{D}_1, \mathcal{D}_2),  \\ \nonumber
P(0,1|1,1) &=& \mathcal{A}_1 F_0(\mathcal{C}_2, \mathcal{C}_1),  \\ \nonumber
P(1,1|1,1) &=& \mathcal{A}_2 F_0(\mathcal{D}_2, \mathcal{D}_1),
\end{eqnarray}
where 
$\mathcal{C}_1 \equiv \alpha \cos{\frac{\theta}{2}}\sin{\frac{\tilde{\theta}}{2}}+ \beta \sin{\frac{\theta}{2}}\cos{\frac{\tilde{\theta}}{2}}$,
$\mathcal{C}_2 \equiv \alpha \cos{\frac{\theta}{2}}\cos{\frac{\tilde{\theta}}{2}}- \beta \sin{\frac{\theta}{2}}\sin{\frac{\tilde{\theta}}{2}}$,
$\mathcal{D}_1 \equiv -\alpha \sin{\frac{\theta}{2}}\sin{\frac{\tilde{\theta}}{2}}+ \beta \cos{\frac{\theta}{2}}\cos{\frac{\tilde{\theta}}{2}}$,
and $\mathcal{D}_2 \equiv \alpha \sin{\frac{\theta}{2}}\cos{\frac{\tilde{\theta}}{2}}+ \beta \cos{\frac{\theta}{2}}\sin{\frac{\tilde{\theta}}{2}}$.
Here only $F_0$ appears since we can represent $F_1(p,q)$ by $F_0(q,p)$ using bases relabeling invariance.

We will study constraints on the functional form of modified quantum probability $F_0(\alpha_0, \alpha_1)$
in Eq. (\ref{eq:modify}) under the no-signaling condition.
The no-signaling condition between Alice and Bob requires that an outcome of Bob's measurement 
must not depend on the choice of measurement axes of Alice. This condition is satisfied when
$\sum_a P(a,b|x,y) = \sum_a P(a,b|\tilde{x},y)$ for all values of $b$ and $y$.
Where $\tilde{x}$ is $1+x$ mod 2.
As a specific example let us consider $b=0$ and $y=0$ case.
In this case the no-signaling condition becomes $P(0,0|0,0)+P(1,0|0,0)=P(0,0|1,0)+P(1,0|1,0)$
which gives
\begin{eqnarray}
\label{eq:nosignal}
&&|\alpha|^2 \\ \nonumber
&=&  \left(|\alpha|^2 \cos^2{\frac{\theta}{2}}+|\beta|^2 \sin^2{\frac{\theta}{2}}\right)
F_0(\alpha \cos{\frac{\theta}{2}}, \beta \sin{\frac{\theta}{2}})
\\ \nonumber &+&
\left(|\alpha|^2 \sin^2{\frac{\theta}{2}}+|\beta|^2 \cos^2{\frac{\theta}{2}}\right)
F_0(\alpha \sin{\frac{\theta}{2}}, \beta \cos{\frac{\theta}{2}})
\end{eqnarray}
The equality cannot be satisfied in general by arbitrary functions $F_0(p,q)$.
Here $\theta$ is the angle of Alice's measurement axis so that Alice can determine her $\theta$
arbitrarily. The theory which restricts on the measurement axis cannot be considered as a proper physical theory.
There are two kinds of solutions of Eq. (\ref{eq:nosignal}) for arbitrary $\theta$.
One is $|\alpha|=|\beta|$ which becomes the Bell state. The post-quantum theories 
must include other states than the Bell state because of the symmetry of a Hilbert space.
A unitary evolution corresponding to a symmetry operation in a Hilbert space will change the Bell state to
other state in general.
 Therefore this solution cannot give a full physical theory.
Another kind can be found by setting $\theta = \pi /2$, then the functional form of $F_0(x,y)$ are determined such as
\begin{eqnarray}
|\alpha|^2 = F_0( \frac{\alpha}{\sqrt{2}}, \frac{\beta}{\sqrt{2}}).
\end{eqnarray}
Considering the state normalization condition this implies $F_0(\alpha, \beta) = |\alpha|^2$ which is the Born rule.
This implies even the minimal modification of the quantum probability rule only for Bob 
is restricted by no-signaling condition.
Therefore, we conclude that we cannot make a post-quantum theory by the modification of the quantum probability rule
from the Born rule without violating no-signaling principle.

\subsection{Simulating the PR boxes}
It is still interesting, however, to consider whether the PR box can be simulated by our model
since the concrete mechanism for generating non-local boxes including the PR box 
is lack except the linear combinations or wiring of two nonlocal boxes \cite{Brunner1}. 

To simulate the non-local boxes we use the following explicit functional form for the probability assignment
\begin{eqnarray}
\label{eq:powerftn}
 F_0(\alpha_0, \alpha_1)= \frac{|\alpha_0|^n}{ |\alpha_0|^n +|\alpha_1|^n}.
\end{eqnarray}
When the initial state $|\psi\rangle_{AB}$ in Eq. (\ref{eq:state}) is the Bell state, i.e., 
for $|\alpha|=|\beta|=1/\sqrt{2}$ the probability distributions $P(a,b|x,y)$ become
\begin{eqnarray}
\label{Eq:PDBell}
P(0,0|0,0) &=& P(1,1|0,0)=\frac{1}{2}, \\ \nonumber
 P(1,0|0,0)&=&P(0,1|0,0) =0, \\ \nonumber
P(0,0|1,0) &=& P(1,1|1,0)=\frac{1}{2}\frac{|\cos{\frac{\theta}{2}}|^n}{|\cos{\frac{\theta}{2}}|^n+|\sin{\frac{\theta}{2}}|^n},
\\ \nonumber
P(1,0|1,0) &=& P(0,1|1,0)=\frac{1}{2}\frac{|\sin{\frac{\theta}{2}}|^n}{|\cos{\frac{\theta}{2}}|^n+|\sin{\frac{\theta}{2}}|^n},
\\ \nonumber
P(0,0|0,1) &=& P(1,1|0,1)=\frac{1}{2}\frac{|\sin{\frac{\tilde{\theta}}{2}}|^n}{|\cos{\frac{\tilde{\theta}}{2}}|^n+|\sin{\frac{\tilde{\theta}}{2}}|^n}, \\ \nonumber
P(1,0|0,1) &=& P(0,1|0,1)=\frac{1}{2}\frac{|\cos{\frac{\tilde{\theta}}{2}}|^n}{|\cos{\frac{\tilde{\theta}}{2}}|^n+|\sin{\frac{\tilde{\theta}}{2}}|^n},
\\ \nonumber
P(0,0|1,1)&=& P(1,1|1,1)= \frac{1}{2} \frac{|\sin{\frac{\theta+\tilde{\theta}}{2}}|^n}
{|\sin{\frac{\theta+\tilde{\theta}}{2}}|^n + |\cos{\frac{\theta+\tilde{\theta}}{2}}|^n} \\ \nonumber
P(1,0|1,1)&=& P(0,1|1,1)= \frac{1}{2} \frac{|\cos{\frac{\theta+\tilde{\theta}}{2}}|^n}
{|\sin{\frac{\theta+\tilde{\theta}}{2}}|^n + |\cos{\frac{\theta+\tilde{\theta}}{2}}|^n}
\end{eqnarray}
One can easily check that the no-signaling condition is satisfied for arbitrary $\theta$, $\tilde{\theta}$, and $n$.
This implies that there is no restriction on the measurement angle for the Bell state.
The condition that the above probability
distributions reproduce the probability distributions of the PR Box requires, in the limit of $n\rightarrow \infty$,
\begin{eqnarray}
\label{Eq:Angle}
&&\Big|\cos{\frac{\theta}{2}}\Big|> \Big|\sin{\frac{\theta}{2}}\Big|,
~~~\Big|\sin{\frac{\tilde{\theta}}{2}}\Big|>\Big|\cos{\frac{\tilde{\theta}}{2}}\Big| \\ \nonumber
\mbox{and } &&
\Big|\cos{\frac{\theta + \tilde{\theta}}{2}}\Big| >\Big|\sin{\frac{\theta + \tilde{\theta}}{2}}\Big|.
\end{eqnarray}
To satisfy the above requirements, the angle $\theta$ must be either in the first or the fourth quadrant and the angle $\tilde{\theta}$ either in the
second or the third quadrant. Additionally,
the angles $\theta$ and $\tilde{\theta}$ also have to satisfy
one of $
0< \theta + \tilde{\theta} <\frac{\pi}{2}$, $\frac{3\pi}{2} < \theta + \tilde{\theta} <\frac{5\pi}{2}$ and
$ \frac{7\pi}{2} <\theta + \tilde{\theta} <4 \pi$.
There is always a suitable $\tilde{\theta}$ for any $\theta$ to satisfy one of those conditions.
For these angles the probabilities go to $1/2$
for inputs and outcomes which satisfy $a \oplus b=xy$ and zeros for others in the limit of $n\rightarrow \infty$. 
These probability distributions are the same as those of the PR box.
Therefore the PR box can be simulated by the Bell states with the modified quantum probability rule in
Eq. (\ref{eq:powerftn}) for $n = \infty$.


\section{Other nonlocal boxes}
The joint probability distributions in Eq. (\ref{Eq:PDBell}) are continuous functions of 
the angles $\theta$, $\tilde{\theta}$ and the power $n$ under
the modified quantum probability rule Eq. (\ref{eq:powerftn}), therefore,
they will describe other nonlocal boxes for other $\theta$, $\tilde{\theta}$ and $n$. 
The joint probability distributions in our model is obtained by the modified quantum probability rule 
for the quantum measurement so that different observable sets described by $\theta$ and $\tilde{\theta}$ 
could generate other joint probability distributions also depending on the power $n$ of the modified Born rule. 
In this way we can get a variety of nonlocal correlations which generate
nonlocal boxes. Another interesting example is the nonlocal box (NLB) which can be generated by the observables
in the standard CHSH inequality \cite{Clauser}.
We will study the nonlocality of these nonlocal boxes in this section.

\subsection{CHSH nonlocality for our non-isotropic NLBs}
First we will study the nonlocality of the non-isotropic NLBs which is described by the 
joint probability distributions in Eq. (\ref{Eq:PDBell}). We will explain what means the non-isotropic boxes later.
We will use the CHSH nonlocality \cite{Forster}
to measure the nonlocality of our non-isotropic NLBs.

The CHSH nonlocality of our nonlocal boxes $P_{NL}$ is defined by the value 
\begin{eqnarray}
\label{Eq:CHSHvalue}
\mathcal{B}(P_{NL}) = \max_{xy}|C_{xy}(P_{NL})+C_{x\tilde{y}}(P_{NL})+C_{\tilde{x}y}(P_{NL})-
C_{\tilde{x}\tilde{y}}(P_{NL})|,
\end{eqnarray}
where $x$ and $y$ are two binary inputs and $\tilde{x}=x+1$ and $\tilde{y}=y+1$ modulo $2$ as before.
The $C_{xy}$ is the correlation function defined as
\begin{eqnarray}
\label{Eq:Correl}
C_{xy}(P_{NL})= P(0,0|x,y)+P(1,1|x,y) -P(0,1|x,y) -P(1,0|x,y).
\end{eqnarray}
To find out the specific inputs $x$ and $y$ which give the maximum value for the CHSH value $\mathcal{B}(P_{NL})$
 we use the continuity of the joint probability functions with respect to the power $n$.
When $n$ goes to infinity, the joint probability distributions becomes those for the PR box
which give the CHSH value 4. This corresponds to $C_{11}(P_{NL}) \rightarrow -1$ for $n \rightarrow \infty$ so
$\tilde{x}$ and $\tilde{y}$ must be 1. That is, $x$ and $y$ are $0$.

\begin{figure}
\label{FIG:CHSHTheta}
\includegraphics{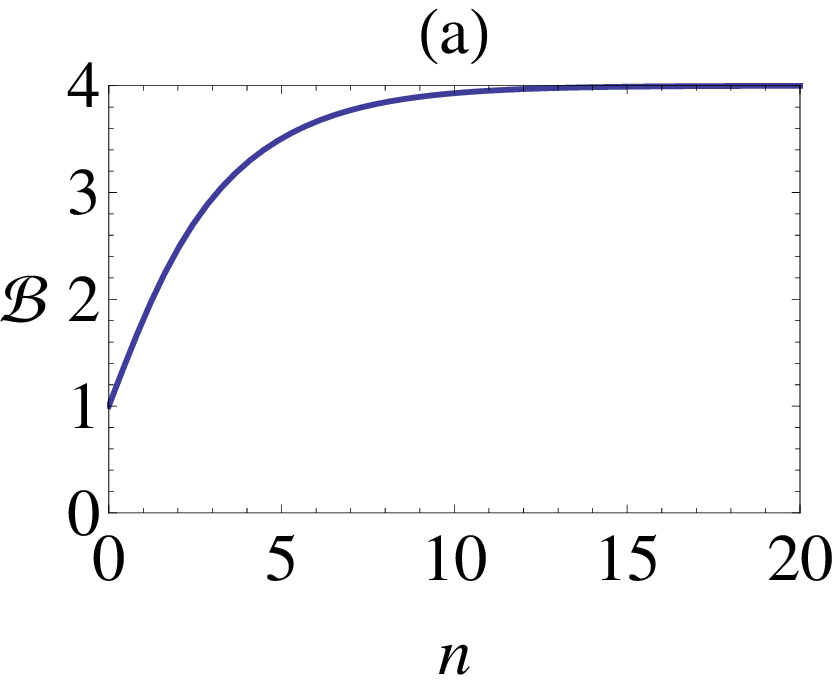}
\includegraphics{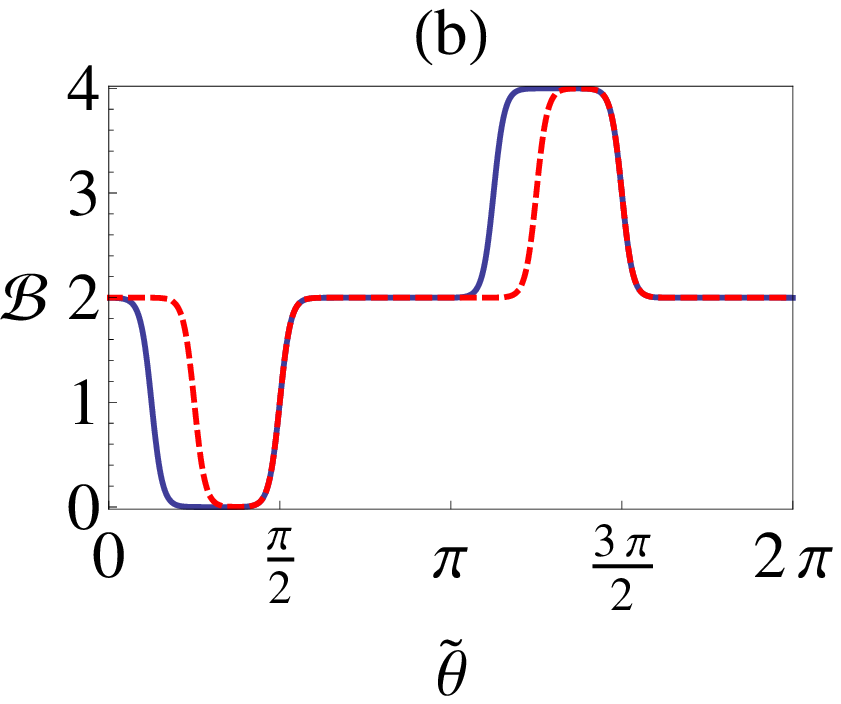}
\vspace{5.2cm}
\caption{(Color online) (a) This shows the dependence of the CHSH value $\mathcal{B}(P_{NL}$ on the power $n$.
(b) This shows the dependence of $CHSH(P)$ on $\tilde{\theta}$ for $n=10$ and
two different $\theta$ values, $\frac{3\pi}{8}$ and $\frac{\pi}{4}$.}
\label{Fig:CHSHDifferentTheta} 
\end{figure}

Therefore the CHSH value of our non-isotropic NLBs for arbitrary $n$ becomes
\begin{eqnarray}
\mathcal{B}(P_{NL})&=& 1+ g_0(\sin{\frac{\tilde{\theta}}{2}},\cos{\frac{\tilde{\theta}}{2}}) - g_1(\sin{\frac{\tilde{\theta}}{2}},\cos{\frac{\tilde{\theta}}{2}}) + 
g_1(\sin{\frac{{\theta}}{2}},\cos{\frac{{\theta}}{2}}) \\ \nonumber
&-& g_0(\sin{\frac{{\theta}}{2}},\cos{\frac{{\theta}}{2}}) +
g_1(\sin{\frac{\theta+\tilde{\theta}}{2}},\cos{\frac{\theta+\tilde{\theta}}{2}}) -
g_0(\sin{\frac{\theta+\tilde{\theta}}{2}},\cos{\frac{\theta+\tilde{\theta}}{2}}),
\end{eqnarray}  
where $g_0(\alpha_0, \alpha_1)= |\alpha_0|^n/(|\alpha_0|^n +|\alpha_1|^n)$ and 
$g_1(\alpha_0, \alpha_1)= |\alpha_1|^n/(|\alpha_0|^n +|\alpha_1|^n)$
The results are shown in the Fig. 2. 
Fig. 2 (a) shows a monotonic increasing behavior of the CHSH value $\mathcal{B}(P_{NL})$ for the power $n$.
The angles of measurement axes $\theta$ and $\tilde{\theta}$ satisfy the condition to simulate the PR box.
In this figure $\theta=\pi/4$ and $\tilde{\theta}=11\pi/8$.
Fig 2 (b) shows the angle $\tilde{\theta}$ dependence of the CHSH value for $n=20$ which is large 
enough for the CHSH value to saturate to its maximum value.
The real line and dashed line corresponds to $\theta=3\pi/8$ and $\theta=\pi/4$ respectively.
There are two kinds of change in the graphs. The first kind of change occurs at different values of $\tilde{\theta}$.
This kind of change occurs first in the real line and is followed by the dashed line. This is because
$\theta+\tilde{\theta}$ of the real line is greater than that of the dashed line. 
The $\theta+\tilde{\theta}$ determines the probability distributions for $x=1$ and $y=1$ as one can see 
in Eq. (\ref{Eq:PDBell}).
The second kind of change start at the same point, i.e., $\tilde{\theta}=\pi/2$ and 
$\tilde{\theta}=3\pi/2$. These values are related with the condition $\tilde{\theta}$ must be
either in the second or the third quadrant, $\pi/2 < \tilde{\theta} <3\pi/2$
 for the CHSH value to be the maximum value for $n\rightarrow \infty$. 



These results imply that the application of the modified quantum probability rule provide the systematic way
of obtaining new nonlocal boxes. These nonlocal boxes are not isotropic since the correlation functions does not
satisfy the condition for the unbiased marginal distributions \cite{Masanes}
\begin{eqnarray}
\label{Eq:Iso}
C_{00}=C_{01}=C_{10}=-C_{11}.
\end{eqnarray}
 The properties of nonlocality such as information causality \cite{Pawlowski} and
computational complexity \cite{Brunner} et al. are studied by the nonlocal boxes 
so the discovery of the systematic way to
obtain new nonlocal boxes is very important.

\subsection{Nonlocal boxes defined by the CHSH observables}
The joint probability distributions in our model depend not only on the state and the probability rule but
the observables. It is interesting to consider the standard CHSH observables and the CHSH inequality for the 
modified quantum probability rule.
The CHSH parameter $\mathcal{B}$ defined as
\begin{eqnarray}
\mathcal{B}= \langle \mathcal{Q}\mathcal{S} \rangle + \langle \mathcal{R} \mathcal{S} \rangle
+\langle \mathcal{R} \mathcal{T}\rangle -\langle \mathcal{Q} \mathcal{T}\rangle.
\end{eqnarray}
Where $\mathcal{Q}= \sigma_z^A$, $\mathcal{S}= -(\sigma_z^B +\sigma_x^B)/\sqrt{2}$,
$R=\sigma_x^A$, and $T=(\sigma_z^B -\sigma_x^B)/\sqrt{2}$ are the CHSH observables. 
The superscripts $A$ and $B$ represent
Alice and Bob respectively. And $\sigma_i$'s are Pauli matrices.
In our case Alice measures first before Bob's measurement.
An expectation value $\langle \mathcal{Q}\mathcal{S}\rangle$ is a quantum correlation of two observables
$\mathcal{Q}$ and $\mathcal{S}$.
Since observables have eigenvalues $\pm 1$,
the expectation value $\langle \mathcal{Q}\mathcal{S}\rangle$ is calculated as follows
\begin{eqnarray}
\langle \mathcal{Q}\mathcal{S}\rangle &=& 1 \times P(1_{\mathcal{Q}}) \left\{ 1 \times P(1_{\mathcal{S}}|1_{\mathcal{Q}})+
(-1)\times P(-1_{\mathcal{S}}|1_{\mathcal{Q}})\right\} \\ \nonumber
&+&
(-1) \times P(-1_{\mathcal{Q}}) \left\{ 1 \times P(1_{\mathcal{S}}|-1_{\mathcal{Q}})+
(-1)\times P(-1_{\mathcal{S}}|-1_{\mathcal{Q}})\right\}.
\end{eqnarray}
$P(\pm1_{\mathcal{Q}})$ are probabilities to obtain eigenvalues $\pm 1$ respectively 
for the observable $\mathcal{Q}$.
To calculate $P(\pm1_{\mathcal{Q}})$ the initial state must be expanded by using the eigenstates of the observable $\mathcal{Q}$.
The probability of outcomes $\pm 1$ for observables $\mathcal{Q}$ and $\mathcal{R}$ of Alice 
are determined by the Born rule in our model.
 After the projective measurement of Alice the state of Bob will be
projected to the pure state in the two-dimensional Hilbert space in which the probability for Bob's measurement
is determined by the modified quantum probability rule.
For example the $P(\pm1_{\mathcal{S}}|\pm1_{\mathcal{Q}})$ are the conditional probability 
calculated by using modified quantum probability rule for the observable $\mathcal{S}$ on 
the projected state of Bob after projective measurement of Alice by the observable $\mathcal{Q}$.

We can define the nonlocal boxes by the CHSH observables with the modified Born rule. 
Let these nonlocal boxes be $P_{CHSH}(n)$, where $n$ represents the dependence on the power 
of the modified Born rule.
The binary inputs $x=0$ and $x=1$ of $P_{CHSH}(n)$ correspond to the measurements 
by $R$ and $Q$ respectively and $y=0$ and $y=1$ to the measurements by $T$ and $S$ respectively.
Then $\langle RT\rangle$ becomes the same as the correlation function $C_{00}(P_{CHSH}(n))$ 
defined in Eq. (\ref{Eq:Correl}).
As a result the CHSH parameter $\mathcal{B}$ also becomes the same as the CHSH value $\mathcal{B}(P_{CHSH}(n))$.
These nonlocal boxes $P_{CHSH}(n)$ are isotropic nonlocal boxes which satisfy the condition 
of the unbiased marginal distributions Eq. (\ref{Eq:Iso}). 

The CHSH parameter is calculated as
\begin{eqnarray}
\mathcal{B}=4\frac{\sqrt{2+\sqrt{2}}^n-\sqrt{2-\sqrt{2}}^n}{\sqrt{2+\sqrt{2}}^n + \sqrt{2-\sqrt{2}}^n}
\end{eqnarray}
Fig. 3 shows the dependence of the CHSH parameter $\mathcal{B}$ on the power $n$. 
When $n$ changes from zero to infinity,
the CHSH $\mathcal{B}$ covers all values up to $4$.
Therefore the Bell state with the modified quantum probability in Eq. (\ref{eq:powerftn})
for arbitrary $n$ can simulate nonlocal boxes with all CHSH $\mathcal{B}$ higher than the quantum upper bound.  
When $n$ goes to infinity the CHSH parameter becomes
the maximum value $4$ as expected for the PR box.
The $\mathcal{B}$ goes to the approximate value $4$ very rapidly so that $\mathcal{B}$ becomes approximately 3.999 for $n=10$.
According to Brassard {\it et al.} \cite{Brassard1}, the isotropic nonlocal boxes with $\mathcal{B}$ more than $4\sqrt{2/3}$
makes communication complexity trivial.
This value $4\sqrt{2/3}$ is achieved for $n\approx 2.601$.

\begin{figure}
\centering
\includegraphics[scale=0.7]{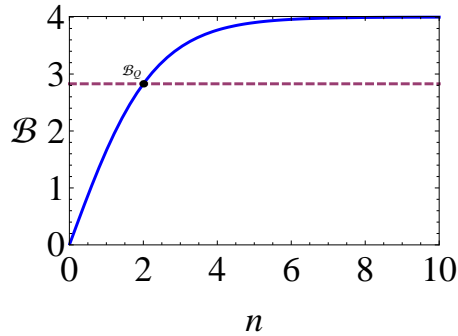}\\
\caption{(Color online) The CHSH parameter $\mathcal{B}$ versus power $n$ of the modified Born rule of Bob. 
It shows $\mathcal{B}$ approaches quickly to
$4$ with increasing $n$. The dot represent the quantum Tsirelson's bound $2\sqrt{2}$. 
The point $\mathcal{B}_Q$ represents the $\mathcal{B}$ for the Born rule of Bob.}
\label{fig:2}
\end{figure}

\section{Discussions and Summary}
The theory to make communication complexity trivial is strongly believed not to exist \cite{Brunner1}.
There was, however, no concrete reason for this belief. 
In this paper we have studied the possibility of the post-quantum theory candidates by applying the modified quantum 
probability rule for the second party under the local measurement requirements. 
The measurement of second party is always on the projected pure separable state so the modification of 
the quantum probability rule is minimal. 
This minimal modification of the quantum probability gives the post-quantum theory candidates which
could have the nonlocality greater than that of the quantum theory.
These post-quantum theory candidates, however, cannot be consistent with the symmetry of the Hilbert space
which requires the freedom on the angle of measurement axis. 
In the post-quantum theory candidates the measurement angle for an arbitrary entangled state is restricted
under the no-signaling condition.
In a linear quantum mechanics Hilbert space description is natural so that we believe the only possible extension
of the quantum mechanics comes from the measurement
postulates, especially, the quantum probability rule. 
Therefore our study suggests that there is no physical theories 
other than the quantum theory under the condition of the no-signaling principle and the local measurement requirements.
The theory with $n\le 2$ is not our concern since the nonlocality of those theories are covered by quantum mechanics.
Moreover, the nonlinear extensions of quantum mechanics proposed by Weinberg \cite{Weinberg}
might be used to send superluminal signals \cite{Weinberg, Gisin1, Polchinski}.
This reinforce our belief that there are no other physical theories than
the quantum mechanics under the condition of no-signaling principle and local measurements.
Therefore this could be a partial explanation for why the nonlocality of the quantum theory is limited.

In summary, we have studied the effect of the modified quantum probability rule for the quantum
measurement on the nonlocality.
The nonlocal quantum correlation manifests itself by correlations of measurement outcomes.
Therefore the modification of the quantum probability rule for the measurement changes nonlocality 
between two spacelike separate
parties. The change of nonlocality was represented by the joint probability distributions.
The joint probability distributions, however, must be given by the Born rule
when the no-signaling condition and the freedom of choice of measurement axes are required.
Other quantum postulates than the quantum measurement postulates are about Hilbert space
which is hard to modify within physically reasonable bound.
We have shown that the probability distributions of the PR box can be approximately simulated 
by our model using explicit function of the modified quantum probability. 
When the power of quantum probability goes to infinity for proper values of the 
measurement angles, the probability distributions of the PR box is reproduced by the Bell state.
We have shown that for the other value of the power and measurement angles the Bell state can simulate 
various nonlocal boxes. We have found the probability distributions for the anisotropic nonlocal boxes 
with the CHSH values up to 4. We have shown the CHSH observables in the usual CHSH inequality generate 
the isotropic nonlocal boxes with the CHSH values up to 4.
This implies that our model can be used as a systematic way to obtain various nonlocal boxes.
The nonlocal boxes with quantum theory-like property will help to understand
the non-locality related properties such as communication complexity and information
causality more deeply.

\section*{Acknowledgments}

 The authors are grateful for good hospitality and helpful discussions in KIAS.
This work was supported by National Research Foundation of Korea Grant funded by the Korean
Government (2011-0005740).


\end{document}